\documentclass[conference]{IEEEtran}
\ifCLASSINFOpdf
  \usepackage[pdftex]{graphicx}
\else
\fi
\usepackage[cmex10]{amsmath}
\usepackage{algorithmic}

\usepackage[numbers]{natbib}
\usepackage{notoccite}
\usepackage{amssymb}
\usepackage{color}
\usepackage{amsmath}

\begin{document}
%
\title{\huge{Graph embeddings of dynamic functional 
connectivity reveal discriminative patterns of task engagement in HCP data}}
\author{\IEEEauthorblockN{Ricardo Monti\IEEEauthorrefmark{1},
Romy Lorenz\IEEEauthorrefmark{2},
Peter Hellyer\IEEEauthorrefmark{2}, 
Robert Leech\IEEEauthorrefmark{2},
Christoforos Anagnostopoulos\IEEEauthorrefmark{1} and
Giovanni Montana\IEEEauthorrefmark{1}\IEEEauthorrefmark{3}}
\IEEEauthorblockA{\IEEEauthorrefmark{1}Department of Mathematics, Imperial College London}
\IEEEauthorblockA{\IEEEauthorrefmark{2}Computational, Cognitive and Clinical Neuroimaging Laboratory, Imperial College
London}
\IEEEauthorblockA{\IEEEauthorrefmark{3}Department of Biomedical Engineering, King’s College London
}}
\maketitle
\begin{abstract}

There is increasing evidence to suggest functional connectivity networks are non-stationary.
This has lead to the development of novel methodologies with which to accurately estimate 
time-varying functional connectivity networks.
Many of these methods provide unprecedented temporal granularity by estimating 
a functional connectivity network at each point in time; 
resulting in high-dimensional output which can be studied in a variety of ways.
One possible method is to employ graph embedding algorithms. 
Such algorithms effectively map 
estimated networks from high-dimensional spaces down to a 
low dimensional vector space; thus facilitating visualization, interpretation and classification. 
In this work,
the dynamic properties of functional connectivity 
are studied using working memory task data from the
Human Connectome Project.
A recently proposed method is employed to estimate dynamic functional connectivity networks. 
The results are subsequently analyzed using two
graph embedding methods based
on linear projections. These methods are shown to 
provide informative embeddings that can be directly 
interpreted as functional connectivity networks.

\end{abstract}
\begin{IEEEkeywords} dynamic networks; Graph embedding; Brain decoding; visualisation; fMRI \end{IEEEkeywords}



%
\IEEEpeerreviewmaketitle

\section{Introduction}

Traditional analysis of fMRI data focused primarily on studying the patterns of individual brain regions.
Recently, there has been a growing interest in understanding relationships between distinct regions; referred to as 
brain connectivity. 
The subsequent study of brain connectivity has provided fundamental insights into the 
human brain \citep{bullmore}. 

A cornerstone in the understanding of brain connectivity is the notion that 
connectivity can be represented as a graph or network composed of a set of nodes interconnected by
a set of edges. 
Nodes can represent anything from individual
voxels to entire regions of interest (ROIs). Here each node
is associated with its own time course of imaging data. This
can subsequently be used to estimate the connections between
nodes, defined as the edge structure of the network. 
In this work we consider estimating edge structure according to 
statistical dependencies between nodes. 
The resulting networks are referred to as 
functional connectivity networks. 

Until recently, functional connectivity networks had been assumed to be stationary. This implied that only a 
global measure of dependency was considered for each pair of nodes. However, there is growing evidence 
to suggest that functional connectivity networks are highly non-stationary \citep{allen2012tracking}. 
This has lead to mounting interest in quantifying dynamic changes in brain connectivity and has 
resulted in the proposal of several novel methodologies \citep{allen2012tracking, leonardi2013principal, monti2014estimating}. 
Such methodologies have provided unprecedented details relating to the restructuring and 
temporal evolution of functional connectivity networks. 
However, understanding and interpreting the high-dimensional output 
provided by such methods has posed novel statistical challenges.
Many of these challenges arise from the fact that while graphs offer a 
rich representation of the data at hand, measuring similarities across 
multiple graphs remains non-trivial. 
One potential method with which to address this issue is 
to employ graph embedding techniques.
Briefly, such methods look
to map estimated graphs or networks down to a
low-dimensional vector space. Such methods have been shown to be well-suited 
to the task of classifying functional networks; 
often referred to as brain decoding
\citep{richiardi2011decoding}. 




The objectives of this work are two fold. First, we present an extensive study of the working memory 
HCP data \citep{elam2014human} using the recently proposed 
Smooth Incremental Graphical Lasso Estimation (SINGLE) algorithm \citep{monti2014estimating}.
Second, the estimated functional connectivity networks are then studied
using two distinct graph embedding algorithms.
The first is based on Principal Component Analysis (PCA) and is strongly related to the eigen-connectivity method
developed in \cite{leonardi2013principal}. This embedding can be interpreted a mapping from
networks to a low-dimensional vector space that captures the maximal variability.
The second approach corresponds to a novel supervised embedding algorithm which is based on Linear
Discriminant Analysis (LDA). 
This algorithm can be interpreted as a mapping to a low-dimensional vector space that maximizes 
discriminatory power across the various tasks or states. 
The strengths of these embeddings is that they are both based on
linear projections. As a result, the transformation described by the 
embeddings can be directly interpreted as a functional connectivity network. 

The remainder of this paper is organised as follows: in Section \ref{methods_sec} we describe two distinct 
graph embedding algorithms. These are applied to HCP data  in Section \ref{ResultSection}.

\section{Methods}
\label{methods_sec}

Throughout this section it is assumed we have access to fMRI time series over a fixed set of $p$ nodes for $S$ subjects.
We write $X^{(s)} \in \mathbb{R}^{n \times p}, s \in \{1, \ldots, S\}$ to denote the 
observed time series for the $s$th subject.
We write $X_i^{(s)} \in \mathbb{R}^{1 \times p}$ to
denote the BOLD measurements at the $i$th observation for $i=1, \ldots, n$.

For each subject, we first look to estimate time-varying functional connectivity
networks at each point in time. This is achieved using the 
recently proposed Smooth Incremental Graphical Lasso Estimation (SINGLE) algorithm \citep{monti2014estimating}.
The resulting estimates over all subjects are then studied using two distinct graph embedding
methods. 
The first method is strongly related to the eigen-connectivities method of \cite{leonardi2013principal}
and is designed to provide a low-dimensional embedding which maximizes explained variance.
The second method is a novel supervised embedding method based on Linear Discriminant Analysis (LDA). 
Here the objective is to obtain a low-dimensional embedding which provides the maximum discriminatory
power between various classes (in the case of HCP data these will be the various tasks undertaken by the subject).

The remainder of this section is organized as follows: in Section \ref{SINGLE} we describe the method used to 
obtain time-varying estimates of functional connectivity networks for each subject independently.
In Sections \ref{PCAembed} and \ref{LDAembed} the two graph embedding methodologies are described.

\subsection{Estimating non-stationary functional connectivity networks}
\label{SINGLE}

In this section, we infer dynamic functional connectivity networks by estimating the corresponding 
precision (inverse covariance) matrix at each observation. Thus, for the $s$th subject we estimate 
a sequence of precision matrices $\{\Theta^{(s)}_i\} = \{ \Theta^{(s)}_1, \ldots, \Theta_n^{(s)} \}$.
Here $\Theta^{(s)}_i \in \mathbb{R}^{p \times p}$ encodes the partial correlation structure at the $i$th observation
for subject $s$. It follows that we can encode $\Theta^{(s)}_i$ as a graph $G^{(s)}_i$ where
the presence of an edge implies a non-zero entry in the corresponding entry of the precision matrix.

The SINGLE algorithm was employed to estimate time-varying precision matrices, $\{\Theta^{(s)}\}$, 
for each of the $S$ subjects independently.
The SINGLE algorithm is able to accurately estimate functional connectivity networks
by enforcing sparsity and temporal homogeneity constraints.
Sparsity 
is introduced in order to
ensure the estimation problem was well-posed as well as to remove spurious edges introduced by noise.
Meanwhile, the  introduction of temporal homogeneity is motivated by a desire to ensure 
changes in functional connectivity are only reported when strongly corroborated by evidence in the data.
The SINGLE algorithm therefore seeks to find a balance between adequately describing the observed
data while satisfying the aforementioned constraints by minimizing the following
convex objective:
\begin{equation}
\label{SINGLE_cost}
 \{\hat \Theta_i^{(s)}\} = \underset{\Theta_i^{(s)}}{\mbox{argmin}} \left  \{f(\{\Theta_i^{(s)}\}) + g_{\lambda_1, \lambda_2}(\{\Theta_i^{(s)}\})  \right \}.
\end{equation}
Here $f(\{\Theta_i^{(s)}\}) = \sum_{i=1}^T  -\mbox{log det }  \Theta_i^{(s)} + \mbox{trace } ( \hat \Sigma_i^{(s)}  \Theta_i^{(s)})$ 
is proportional to the 
sum of negative log-likelihoods where $\hat \Sigma_i^{(s)}$ is the estimated covariance at time $i$.
The penalty terms are enforced by the second term:
$$g_{\lambda_1, \lambda_2}(\{\Theta_i^{(s)} \}) = \lambda_1 \sum_{i=1}^T || \Theta_i^{(s)}||_1 + \lambda_2 \sum_{i=2}^T || \Theta_i^{(s)} -  \Theta_{i-1}^{(s)}||_1. $$
Regularization
parameters $\lambda_1$ and $\lambda_2$ each determine the extent of sparsity and temporal homogeneity respectively 
and can be tuned in a data-driven manner.


 
\subsection{Graph embedding} 

Having obtained functional connectivity estimates, $\{ \Theta^{(s)}_i\}$ for all $S$ subjects 
we turn to the problem of graph embedding. 
Graph embeddings methods allow us to represent graphs or networks in (potentially low-dimensional) vector spaces. 
Such methods are beneficial as there is no ``standard'' method to measure distances 
between graphs \citep{richiardi2010vector} while vectors are easily interpretable and
can be studied in a variety of ways. Moreover, as we will discuss below, the embeddings
studied here consist of linear projections and therefore yield a unique 
interpretation in terms of 
functional connectivity networks.

We begin
by first calculating the Laplacian of each estimated functional connectivity network:
\begin{equation}
 L_i^{(s)} = (D^{(s)}_i)^{-\frac{1}{2}} (D^{(s)}_i - \Theta^{(s)}_i ) (D^{(s)}_i)^{-\frac{1}{2}},
\end{equation}
where $D^{(s)}_i$ is a diagonal matrix containing the variance of each node respectively.
It therefore follows that each $L^{(s)}_i$ is fully characterized the its $\binom{p}{2}$ upper-triangular entries. 
In the following sections we will use $\{L^{(s)}_i\}$ directly to perform graph embedding.
We define
\begin{equation}
 \mbox{vec}(L^{(s)}) =  \mathbb{R}^{n \times \binom{p}{2}}
\end{equation}
as a matrix where the $i$th row corresponds to the vectorized upper-triangular entries of
the Laplacian 
at the $i$th observation.
The matrix, $L$, consisting of all vectorized Laplacians across all subjects can subsequently
be defined as:
\begin{equation}
 L  = \left [ \mbox{vec}(L^{(1)})^T, \ldots \mbox{vec}(L^{(S)})^T \right ]^T \in \mathbb{R}^{ S \cdot n \times \binom{p}{2}}.
\end{equation}
This process is described in Figure [\ref{TheFig}a]. 
It 
follows that each column of $L$ corresponds directly to one of the $\binom{p}{2}$ possible edges.
As both embeddings studied here consist of linear projections of $L$ onto lower-dimensional subspaces,
they can each be understood as a 
a linear combination of edges and interpreted as 
functional connectivity networks.

\subsubsection{Unsupervised PCA-driven embedding}
\label{PCAembed}

Here we look to obtain a low-dimensional embedding that maximizes the amount of explained variance.
Following from the method described in \cite{leonardi2013principal}, we 
look to achieve this by applying Principal Component Analysis (PCA) to $L$. 
This will yield the linear combination of edges that best summarize the variability in 
functional connectivity networks over time. 

Formally, PCA is an unsupervised dimensionality reduction technique which produces a 
new set of uncorrelated variables. This is 
achieved by considering the $k$ leading eigenvectors of the covariance matrix $ L^T L$, 
defined as the principal components $P_k \in \mathbb{R}^{k \times \binom{p}{2}}$.

The principal components, $P_k$, can be studied in two ways. 
By considering the entries of each principal component we 
are able to quantify the contribution of the corresponding edges. 
Edges which vary highly with a dataset
can therefore be expected to provide a large contribution to the leading principal 
components. 
Moreover, the embedding produced by $P_k$ is obtained as:
\begin{equation}
 P_k \cdot \mbox{vec}(L^{(s)}) \in \mathbb{R}^{k \times n}.
\end{equation}
This yields a $k$-dimensional graph embedding for each subject at each of the $n$ observations.

%


\subsubsection{Supervised LDA-driven embedding}
\label{LDAembed}
While the PCA-driven embedding was 
motivated by understanding the components of
functional connectivity which
demonstrated the greatest variability,
we may also be interested in understanding which functional networks are 
most discriminative across multiple tasks. As a result, a supervised learning approach is taken here. 

We propose the use of LDA to learn the functional connectivity networks which are
most discriminative between tasks. 
LDA is a simple and robust classification algorithm, 
but more importantly,
LDA can also be interpreted as a linear projection. 
As a result, LDA reports the linear combination of edges which are most discriminative 
between tasks. These can subsequently be interpreted as 
a discriminative embedding which reports changes in functional
connectivity induced by a given task.



In such a high-dimensional supervised learning problems it is of paramount importance to avoid overfitting.
Two popular methods to guard against overfitting involve the introduction of
regularization, 
thereby penalizing overly complex models which are more
susceptible to overfitting,
and cross-validation.
Here a combination of both approaches was employed. 
First a variable screening procedure was applied, reducing the
number of candidate variables (i.e., edges) to $p' << \binom{p}{2}$.
This serves to greatly reduce the risk of overfitting as well as 
yield a sparse embedding which is easily interpretable.
The remaining $p'$ selected edges are subsequently used to train 
an LDA classifier. 
Such a classifier will learn the linear projection of selected edges 
which is most discriminative across tasks. This projection will serve 
as our LDA-driven embedding.


%



The screening method employed in this work selected the most reproducible 
edges across all $S$ subjects. This was achieved by fitting an independent 
LDA classifier for the data of each subject. Due to the limited observations per subject, regularization was introduced in the form of an $l_1$ penalty. 
As a result, an $l_1$ penalized LDA model was estimated for each subject.
Such models can be estimated efficiently as described in \citep{clemmensen2011sparse}
and provide the additional benefit of performing variable selection. 
It follows that the regularization parameter will play a fundamental role in 
the variable selection procedure and must therefore be carefully
tuned. 
This issue was addressed via the use of a cross-validation scheme.
The variables which were consistently
selected across all subjects where retained and all others discarded.

Such a screening approach can be interpreted as performing stability selection, as 
described in \citep{meinshausen2010stability}, where the sub-sampling 
is performed by studying each subject independently. 
This serves to discard a large number of 
noisy and non-informative variables, yielding a
Laplacian matrix, $L' \in \mathbb{R}^{S \cdot n \times p'}$,
consisting of only selected variables which have 
demonstrated reproducible discriminative power 
across all subjects.



\section{Materials and Results}
\label{ResultSection}

Working Memory (WM) task data from the Human Connectome Project \cite{elam2014human} was studied with
$S=206$ of the 500 available subjects selected at random. 
During the tasks subjects were presented with blocks of trails
consisting of either 0-back or 2-back WM tasks.
For each subject both a LR and a RL acquisition dataset was studied, however, they were treated as
separate scans and studied separately throughout. Thus 
a total of $2 \times 206=412$ datasets were studied.
Preprocessing involved regression of Friston’s 24 motion parameters from the fMRI data.
Sixty-eight cortical and 16 subcortical ROIs were derived from 
the Desikan-Killiany atlas and the ASEG atlas,
respectively. Mean BOLD timeseries for each of these 84 ROIs were extracted and further 
cleaned by regressing out timeseries sampled from white matter and cerebrospinal fluid. 
Finally, the extracted timecourses were high-pass filtering using a cut-off frequency of $\frac{1}{130}$ Hz.

\subsection{PCA-driven embedding results}

The objective of PCA-driven embedding is to provide a low-dimensional embedding 
which captures a large portion of the variability present in the data.
This was achieved in an unsupervised manner by considering 
the leading $k=2$ principal components of Laplacian matrix $L$. 
The leading $k=2$ components were considered 
here as the emphasis lay on visualization, however further components
could also be studied.
In this section, both the LR and RL acquisitions for each subject were 
considered simultaneously as the goal was to understand variability across 
the entire population. 

Figure [\ref{TheFig}b] shows the average 
embeddings across all $S$ subjects\footnote{Note: only LR acquisition datasets
plotted here, as the task design varied from LR to RL acquisitions.}.
The background is colored to denote the task taking place at each point in time; 
green is used to denote 2-back WM task
while purple denotes a 0-back WM task and a white background is indicative of rest.
We note that in the case of both embeddings
there is a clear oscillatory pattern which is 
strongly correlated with the task. Moreover, the two embeddings are lagged, 
with the second principal component embedding peaking immediately after
the first. 



The functional connectivity networks associated with these embeddings are 
shown in Figures [\ref{TheFig}c]. 
These networks have
been thresholded to retain only 2\% of edges with the largest absolute value.
It can be seen that they reflect independent networks dynamics.
The first component appears to reflect stronger inter-hemispheric coupling, particularly between
motor regions.
The second
component appears to reflect 
increased intra-hemispheric long-range connections between frontal and parietal regions.

\subsection{LDA-driven embedding results}

The motivation behind the use of LDA-driven embedding is to provide an
interpretable embedding which is highly discriminative across 
various tasks. In this section, we studied the contrast between
the 0-back and 2-back WM tasks. 

As noted previously, two datasets where available for each subject. In such a 
supervised learning task it is important to differentiate between 
the LR and RL acquisition datasets as they each employed a distinct task-design.
The approach taken here was to build an LDA-driven embedding using only the LR
acquisition datasets across all subjects and then validate this model using 
the unseen RL acquisition datasets. 
All $\binom{p}{2}$ potential edges were screened as described previously and 
only those selected over 60\% of the time were studied. This reduced 
the number of candidate edges to $p'=126 << \binom{p}{2}$.

\begin{figure}[h]
\centering
\fbox{\includegraphics[width=.4\paperwidth, height=.65\paperheight]{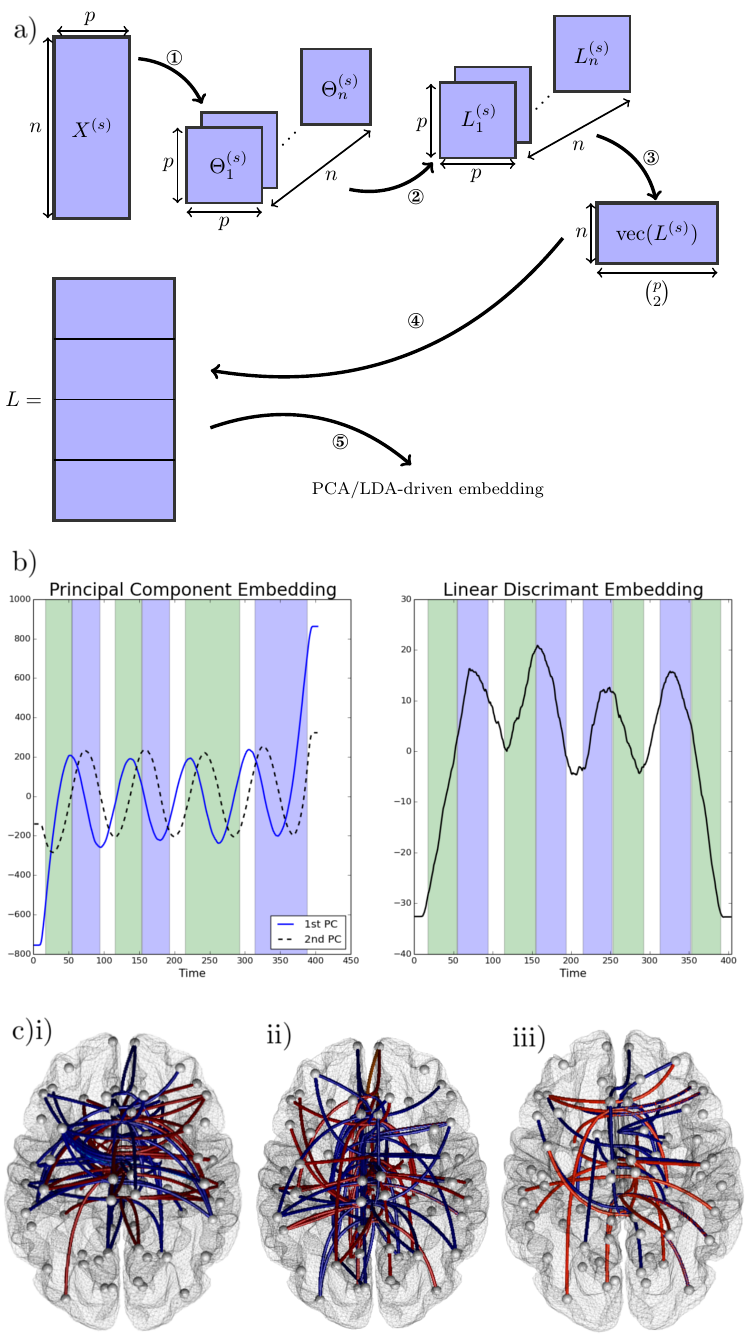}}
\caption{\textbf{a)} The various steps involved in the proposed embedding method are visualized: 1) the SINGLE algorithm is used
to obtain estimates of time-varying precision matrices. 2) The precision matrices are transformed to Laplacian matrices.
3) The Laplacian matrices are vectorized by taking their upper-triangular components.
4) The vectorized Laplacians of all subjects are stacked vertically. 
5) Finally the PCA/LDA-driven embeddings are estimated.
\textbf{b)} The results for the PCA-driven embeddings are plotted for the LR acquisition datasets. We note there is a 
clear periodicity which is strongly correlated with task onset. 
The results for LDA-driven embeddings are also shown for unseen RL acquisition datasets.
We note these are strongly correlated with 0-back
task onset.
\textbf{c)} Functional connectivity networks associated with various embeddings 
(negative edges shown in red, positive in blue)
\textbf{i)} first principal component embedding,
\textbf{ii)} second principal component embedding and \textbf{iii)} LDA-driven embedding.
In the case of the PCA-embeddings the networks where thresholded to leave only 2\% of edges with largest absolute value.
}
\label{TheFig}
\end{figure}

Figure [\ref{TheFig}b] shows the results of applying 
the LDA-driven embedding to the unseen RL acquisition datasets, averaged 
across all $S$ subjects. We note 
the resulting embedding is strongly correlated with the onset of the 
0-back WM task (denoted by the purple background).
This serves as evidence that the estimated LDA-driven embedding is 
able to discriminate between the two WM tasks in a validation dataset.

The corresponding functional connectivity network 
associated estimate LDA-driven embedding is shown in Figure [\ref{TheFig}c]. 
%
We note that the higher loading task condition (2-back WM task; characterized by red edges in Figure [\ref{TheFig}ciii])
is associated with stronger long-range inter-hemispheric connections. 


\section{Conclusion}


We present results from applying two graph embedding methods to HCP working memory task data for 
206 subjects. Both of the embedding methods studied here are based on 
linear projections and therefore yield embeddings that 
are easily interpretable in terms of functional connectivity networks.

The first method studied here is closely related to the eigen-connectivities method
introduced in \citep{leonardi2013principal}. Here PCA is employed to 
capture variability throughout a task across a population of subjects. 
The second graph embedding method studied corresponds to a novel supervised embedding 
algorithm. First, a screening step drastically reduces the total number of potential 
in order to yield interpretable embeddings and  avoid overfitting.
LDA is subsequently used to obtain an embedding that is  discriminative across tasks.

In future, such approaches could be applied to the full range of 
task data provided by the HCP. 
In addition, we note that the 
SINGLE algorithm requires
the input of two regularization parameters which will ultimately affect
the results of the approaches discussed. 
Future work would involve studying the sensitivity of each method to the choice of such parameters.

\section*{acknowledgement}
The data was provided by the Human Connectome Project.



%

{\footnotesize
\bibliographystyle{unsrtnat}
\bibliography{ref}

\begin{thebibliography}{9}
\providecommand{\natexlab}[1]{#1}
\providecommand{\url}[1]{\texttt{#1}}
\expandafter\ifx\csname urlstyle\endcsname\relax
  \providecommand{\doi}[1]{doi: #1}\else
  \providecommand{\doi}{doi: \begingroup \urlstyle{rm}\Url}\fi

\bibitem[Bullmore and Sporns(2009)]{bullmore}
E.~Bullmore and O.~Sporns.
\newblock Complex brain networks: graph theoretical analysis of structural and
  functional systems.
\newblock \emph{Nature}, 10:\penalty0 186--198, 2009.

\bibitem[Allen et~al.(2012)Allen, Damaraju, Plis, Erhardt, Eichele, and
  Calhoun]{allen2012tracking}
E.~Allen, E.~Damaraju, S.~Plis, E.~Erhardt, T.~Eichele, and V.~Calhoun.
\newblock Tracking whole-brain connectivity dynamics in the resting state.
\newblock \emph{Cerebral cortex}, page bhs352, 2012.

\bibitem[Leonardi et~al.(2013)Leonardi, Richiardi, Gschwind, Simioni, Annoni,
  Schluep, Vuilleumier, and Van De~Ville]{leonardi2013principal}
N.~Leonardi, J.~Richiardi, M.~Gschwind, S.~Simioni, J-M. Annoni, M.~Schluep,
  P.~Vuilleumier, and D.~Van De~Ville.
\newblock Principal components of functional connectivity: A new approach to
  study dynamic brain connectivity during rest.
\newblock \emph{NeuroImage}, 83:\penalty0 937--950, 2013.

\bibitem[Monti et~al.(2014)Monti, Hellyer, Sharp, Leech, Anagnostopoulos, and
  Montana]{monti2014estimating}
R.~P. Monti, P.~Hellyer, D.~Sharp, R.~Leech, C.~Anagnostopoulos, and
  G.~Montana.
\newblock Estimating time-varying brain connectivity networks from functional
  mri time series.
\newblock \emph{NeuroImage}, 103:\penalty0 427--443, 2014.

\bibitem[Richiardi et~al.(2011)Richiardi, Eryilmaz, Schwartz, Vuilleumier, and
  Van De~Ville]{richiardi2011decoding}
J.~Richiardi, H.~Eryilmaz, S.~Schwartz, P.~Vuilleumier, and D.~Van De~Ville.
\newblock Decoding brain states from f{M}{R}{I} connectivity graphs.
\newblock \emph{NeuroImage}, 56\penalty0 (2):\penalty0 616--626, 2011.

\bibitem[Elam and Van~Essen(2014)]{elam2014human}
J.~S. Elam and D.~Van~Essen.
\newblock Human connectome project.
\newblock In \emph{Encyclopedia of Computational Neuroscience}, pages 1--4.
  Springer, 2014.

\bibitem[Richiardi et~al.(2010)Richiardi, Van De~Ville, Riesen, and
  Bunke]{richiardi2010vector}
J.~Richiardi, D.~Van De~Ville, K.~Riesen, and H.~Bunke.
\newblock Vector space embedding of undirected graphs with fixed-cardinality
  vertex sequences for classification.
\newblock In \emph{ICPR}, pages 902--905. IEEE, 2010.

\bibitem[Clemmensen et~al.(2011)Clemmensen, Hastie, Witten, and
  Ersb{\o}ll]{clemmensen2011sparse}
L.~Clemmensen, T.~Hastie, D.~Witten, and B.~Ersb{\o}ll.
\newblock Sparse discriminant analysis.
\newblock \emph{Technometrics}, 53\penalty0 (4), 2011.

\bibitem[Meinshausen and B{\"u}hlmann(2010)]{meinshausen2010stability}
N.~Meinshausen and P.~B{\"u}hlmann.
\newblock Stability selection.
\newblock \emph{Journal of the Royal Statistical Society: Series B},
  72\penalty0 (4):\penalty0 417--473, 2010.

\end{thebibliography}
}

\end{document}